\documentclass[seceq]{ptptex}

\usepackage{graphicx}
\usepackage{bm}

\title{
Electromagnetic Mean Squared Radii of $\Lambda (1405)$ 
in Meson-baryon Dynamics with Chiral Symmetry%
}

\author{
Takayasu \textsc{Sekihara},$^{1,}$\footnote{E-mail: 
sekihara@ruby.scphys.kyoto-u.ac.jp}  Tetsuo \textsc{Hyodo}$^{2,3}$ and 
Daisuke \textsc{Jido}$^{2}$%
}

\inst{
$^{1}$Department of Physics, Kyoto University, Kyoto 606-8502, Japan

$^{2}$Yukawa Institute for Theoretical Physics, Kyoto University, 
Kyoto 606-8502, Japan

$^{3}$Physik-Department, Technische Universit\"{a}t M\"{u}nchen, 
D-85747 Garching, Germany
}

\def\be{\begin{equation}}
\def\ee{\end{equation}}
\def\bc{\begin{center}}
\def\ec{\end{center}}
\def\trm{\textrm}
\def\GE{G_{\trm{E}}}
\def\GM{G_{\trm{M}}}
\def\EMSR{\langle r^{2} \rangle _{\trm{E}}}
\def\MMSR{\langle r^{2} \rangle _{\trm{M}}}
\def\KbarN{\bar{K} N}
\def\Kmp{K^{-} p}

\abst{
Electromagnetic mean squared radii of $\Lambda (1405)$ are evaluated 
in chiral unitary approach. In this approach we regard $\Lambda (1405)$ 
as dynamically generated resonances in the octet meson and octet baryon 
scattering, and also as $\KbarN$ bound 
state. Especially for the $\Lambda (1405)$ as $\KbarN$ bound state we 
obtain negative electric mean squared radius. With the small binding 
energy of $\Lambda (1405)$ in the chiral unitary approach, 
our results imply that $\Lambda (1405)$ has structure 
that $K^{-}$ is widely spread around $p$. }

\begin{document}

\maketitle

\section{Introduction}

One of the most interseting physics in hot and/or dense QCD is 
restoration of chiral symmetry. In order to verify the symmetry 
restoration in hot and/or dense system and extract some properties 
of QCD vacuum from the symmetry restoration, many studies have 
been extensively done from both experimental and theoretical sides. 
Among these studies, mesic nuclei in which Nambu-Goldstone boson is 
bound in nuclei are considered as a good subject of the study because 
Nambu-Goldstone bosons are strongly affected by QCD vacuum in finite 
density. Especially kaonic nuclei are interesting because kaon is 
expected to be bound in nuclei due to the strong attractive 
interaction between kaon and nucleon required by chiral symmetry. 

It is suggested that binding energy of $\KbarN$ system in the 
$\Lambda (1405)$ plays an important role for the kaonic 
nuclei.~\cite{Akaishi:2002bg,Yamazaki:2007cs,Hyodo:2007jq,Dote:2008in} 
Further the 
structure of the $\KbarN$ system might be essential to the properties 
of kaonic nuclei. 
Related to the kaonic nuclei, we explain our study of electromagnetic 
mean squared radii of $\Lambda (1405)$ in chiral unitary approach, 
in which $\Lambda (1405)$ is described by meson-baryon dynamics based 
on chiral 
symmetry.~\cite{Kaiser:1995eg,Oset:1997it,Oller:2000fj,Lutz:2001yb,Jido:2003cb}  

In chiral unitary approach, the scattering amplitude is 
obtained by an argeblaic equation:
\be
T_{ij} (\sqrt{s}) = V_{ij} (\sqrt{s}) 
+ \sum _{k} V_{ik} (\sqrt{s}) G_{k} (\sqrt{s}) T_{kj} (\sqrt{s}) ,
\label{eq:Tamp}\ee
with an $s$-wave interaction kernel $V$ given by the 
lowest order chiral perturbation
theory, that is the Weinberg-Tomozawa interaction, and the meson-baryon
loop function $G$. Both $V$ and $G$ are functions of the 
center-of-mass energy 
of the meson-baryon system, $\sqrt{s}$, in matrix forms of 
the meson-baryon channels of strangeness $-1$ and charge $0$. 
This approach has reproduced well the scattering cross 
sections of $K^{-}p$ to
various channels and the mass spectrum of the $\Lambda (1405)$ 
resonance below the $\KbarN$ threshold, giving two states for 
the $\Lambda (1405)$ as poles of the scattering amplitudes
in the complex energy plane, 
($z_{1}=1390-66i \, \trm{MeV}$) and 
($z_{2}=1426-17i \, \trm{MeV}$).~\cite{Oller:2000fj,Jido:2003cb}

\section{Form Factors and Mean Squared Radii of Excited Baryons}

In this section, we discuss the formulation to evaluate the electromagnetic 
form factors and the mean squared radii of excited baryons described 
by the amplitudes obtained from Eq.~(\ref{eq:Tamp}). 
First of all, let us define the electromagnetic form factors of an 
excited baryon with spin $1/2$, 
$|H^{\ast}\rangle$, as matrix elements of the electromagnetic current 
$J_{\trm{EM}}^{\mu}$ in the Breit frame:~\cite{Jido:2002yz,Sekihara:2008qk} 
\be
\left \langle H^{\ast} \left | J_{\trm{EM}}^{\mu} \right | H^{\ast}
\right \rangle _{\trm{Breit}} \equiv \left( \GE (Q^{2}) , \, \GM (Q^{2}) 
\frac{i \bm{\sigma} \times \bm{q}}{2 M_{\trm{p}}} \right) , 
\ee
with the electric and magnetic form factors, $\GE (Q^{2})$ and 
$\GM (Q^{2})$, the 
virtual photon momentum $q^{\mu}$, $Q^{2}=-q^{2}$ and the Pauli 
matrices $\sigma ^{a} \, (a=1, \, 2, \, 3)$. 
The magnetic form factor $\GM (Q^{2})$ is normalized as the 
nuclear magneton $\mu _{\trm{N}} = e / (2 M_{\trm{p}})$ with the 
proton charge $e$ and mass $M_{\trm{p}}$. From these form factors, 
electromagnetic mean squared radii, $\EMSR$ and $\MMSR$, are 
calculated by
\be
\EMSR \equiv - 6 \left . 
\frac{d \GE}{d Q^{2}} \right | _{Q^{2}=0}, \quad 
\MMSR \equiv 
- \frac{6}{G_{\trm{M}}(0)} \frac{d G_{\trm{M}}}
{d Q^{2}} \bigg |_{Q^{2}=0} .
\ee

\begin{figure}[t]
 \bc
 \begin{tabular}{ccc}
    \includegraphics[scale=0.125]{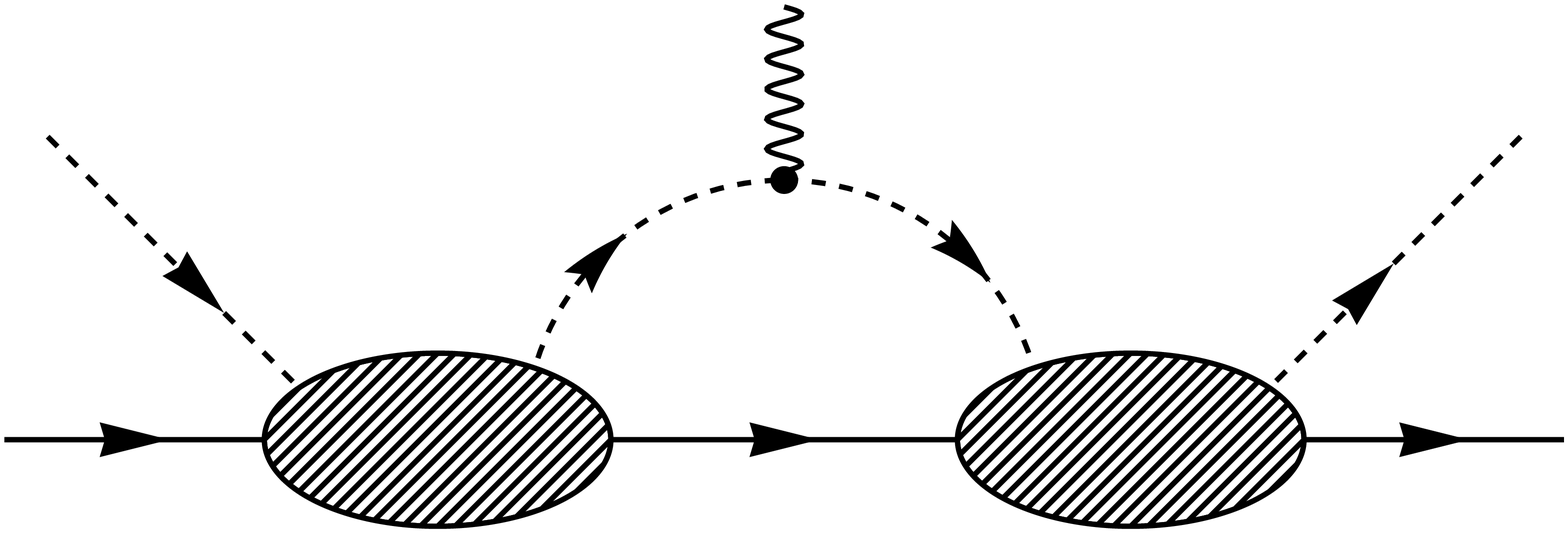} &
    \includegraphics[scale=0.125]{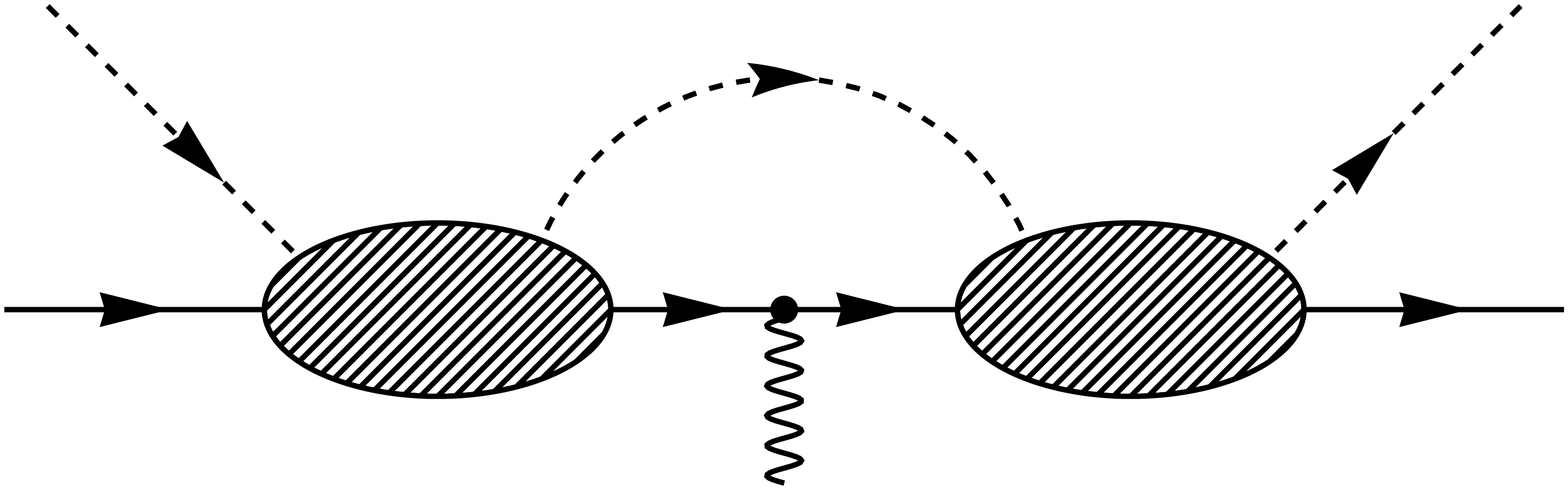} &
    \includegraphics[scale=0.125]{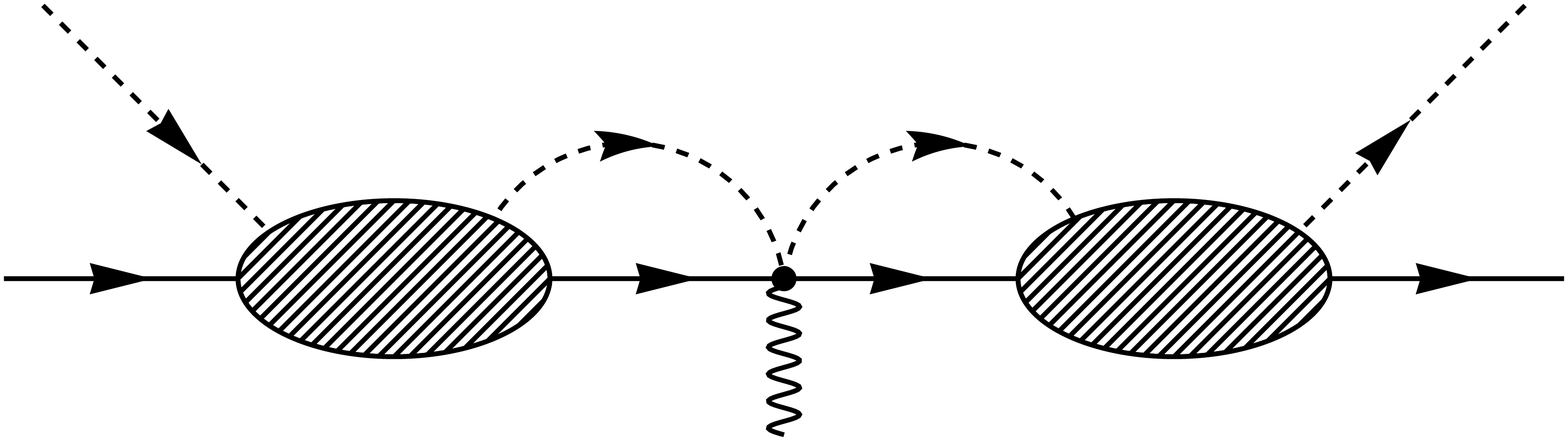} \\
    $- i T_{\gamma 1}^{\mu}$ &
    $-iT_{\gamma 2}^{\mu}$ &
    $-iT_{\gamma 3}^{\mu}$
 \end{tabular}
 \ec
 \caption{Diagrams for the form factor of the $\Lambda(1405)$. 
The shaded ellipses represent the meson-baryon scattering amplitudes.
 \label{diagram}}
\end{figure}

The matrix elements of $J_{\trm{EM}}^{\mu}$ are related to the residue 
of the double pole of the $MB \gamma ^{\ast} \rightarrow MB$ amplitude 
$T^{\mu}_{\gamma ij}$.~\cite{Jido:2002yz,Sekihara:2008qk}  
In our approach, the amplitude $T_{\gamma ij}^{\mu}$ is calculated 
under the assumption that the photon interacts with the $\Lambda(1405)$ 
through the photon couplings to the constituent meson and baryon
of the $\Lambda(1405)$. 
The calculation of the photon coupling should be performed in a gauge
invariant way, since the gauge invariance guarantees to give 
the correct electric charge to the excited baryons, 
$\GE(Q^{2}=0)=Q_{\trm{H}}$. Within the framework of the gauge invariant 
calculation proposed by Ref.~\citen{Borasoy:2005zg}, 
we find the relevant diagrams for our purpose, which have the double 
poles, as shown in Fig.~\ref{diagram}. The photon couplings to the 
mesons and baryons are given by gauging the kinetic terms 
and the effective interaction. Calculating these three 
diagrams and summing up them,
we obtain the $MB \gamma ^{\ast} \rightarrow MB$ amplitude, 
\be
T_{\gamma ij}^{\mu} \equiv T_{\gamma 1 ij}^{\mu} + 
T_{\gamma 2 ij}^{\mu} + T_{\gamma 3 ij}^{\mu} .
\ee
The detailed calculation and the proof of gauge invariance 
of form factors in our scheme is given in Ref.~\citen{Sekihara:2008qk}.

In the present calculation, we did not introduce the form factors 
for the ground state mesons and baryons and treat them as point
particles, because we are interested in the sizes of the excited 
baryon generated by meson-baryon dynamics and in estimation of 
pure dynamical effects. In this formulation, however, inclusion of the 
form factors for the mesons and baryons is straightforward; we 
simply multiply the meson and baryon form factors to each vertex. 
It is worth noting that the electric mean squared radii 
for neutral excited baryons do {\it not} depend on the inclusion 
of the meson and baryon form factor, due to the neutrality of the 
excited baryons.~\cite{Sekihara:2008qk}

\section{Results}

First of all, we briefly mention the results of electromagnetic 
mean squared radii of two states of $\Lambda (1405)$, $z_{1}$ 
and $z_{2}$. As is well known, mean squared radii of resonant 
states, which have decay width, are in general complex. We 
obtain complex mean squared radii of $\Lambda (1405)$. The mean 
squared radii are second moments of the form factor. Taking the 
absolute value, we obtain the second moment for the higher 
$\Lambda (1405)$ state $z_{2}$ as $0.32 \, \trm{fm}^{2}$, which is 
larger than that of neutron $\sim -0.12 \, \trm{fm}^{2}$. This 
means that the $\Lambda (1405)$ has a softer form factor than 
the neutron. 
The detailed discussions about the complex mean squared radii 
are given in Ref.~\citen{Sekihara:2008qk}. 

In order to extract the information of the electromagnetic sizes, 
we perform the following analysis. If the decay width 
of the resonance is small, the mean squared 
radii are close to real numbers. For an estimate of the size, we 
consider $\Lambda (1405)$ as a $\KbarN$ bound state, 
which is generated only by the attractive interaction of the $\KbarN$ 
channel~\cite{Akaishi:2002bg,Yamazaki:2007cs,Hyodo:2007jq} and considered to be a 
origin of higher $\Lambda (1405)$ state $z_{2}$.

Using the same parameter as in the fully coupled-channel case, we 
obtain 
$\EMSR = -2.19 \, \trm{fm}^{2}$ and $\MMSR = 1.97 \, \trm{fm}^{2}$ 
for the $\KbarN$ bound state with mass $1429 \, \trm{MeV}$. 
The negative sign for the electric mean squared radius implies that 
the $K^{-}$ is surrounding around the proton. In addition, with the 
fact that the electromagnetic size of the proton is roughly 
$0.9 \, \trm{fm}$, our result of the electric root mean squared 
radius $\sqrt{| \EMSR |} \simeq 1.48 \, \trm{fm}$ implies 
that $\Lambda (1405)$ has structure of widely spread $K^{-}$ 
clouds around the core of proton with larger size than that of typical 
ground state baryons. 

\begin{figure}[t]
\bc
 \includegraphics[scale=0.3,angle=-90]{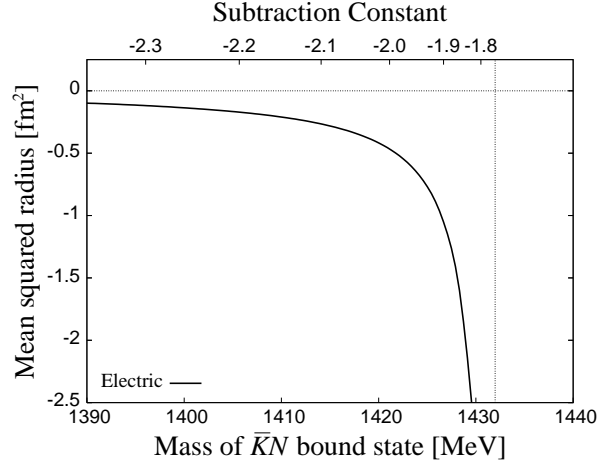}
 \caption{Electric mean squared radius of the $\KbarN$ bound 
state as a function of the mass. The upper horizontal axis denotes 
the corresponding values of the subtraction constants, which 
is a parameter in chiral unitary approach. 
The vertical dotted line represents position of the $\Kmp$ threshold.}
\label{fig:RvsB}
\ec
\end{figure}

In Fig.~\ref{fig:RvsB}, we show the electric mean squared 
radius as a function of the mass of the $\KbarN$ bound state. 
The mass of the $\KbarN$ bound state is controlled by a subtraction 
constant, which is the only one parameter in our approach appearing 
in the loop integral $G$. As one can see, our result is consistent 
with the expectation that the deeper bound states have the smaller 
radii. 

For deep binding energy (about more than $40 \, \trm{MeV}$), 
electric mean squared radius is as small as that of neutron (about less than 
$-0.1 \, \trm{fm}^{2}$). This implies that overlap between kaon and 
nucleon is large, therefore picture of meson-baryon bound state for 
$\Lambda (1405)$ may be destroyed.~\cite{KanadaEn'yo:2008wm} 
For small binding energy (about 
a few $\trm{MeV}$), on the other hand, absolute value of electric 
mean squared radius is about larger than $1 \, \trm{fm}^{2}$, which 
exceeds the typical size of ground state baryons, and overlap between 
kaon and nucleon is small. In the chiral unitary approach, 
we have $\Lambda (1405)$ with about a few $\trm{MeV}$ binding energy 
as the pole position of the higher state $z_{2}$. 
Therefore, our result shown in Fig.~\ref{fig:RvsB} is consistent 
with the picture that $\Lambda(1405)$ is generated by $\KbarN$ 
meson-baryon dynamics.

\section{Summary}

We have calculated electromagnetic mean squared radii of $\Lambda (1405)$ 
in the chiral unitary approach. Calculations are performed in two ways: 
fully coupled-channel approach and approach of $\KbarN$ bound state 
by neglecting all the coupling of $\KbarN$ to the other channels. We 
obtain large negative value ($\lesssim -1 \, \trm{fm}^{2}$) for electric mean 
squared radius for $\KbarN$ bound state with small binding energy 
(about a few $\trm{MeV}$). Furthermore, our result is consistent with 
the picture that $\Lambda (1405)$ is generated by $\KbarN$ meson-baryon 
dynamics. 


%

\end{document}